\documentclass[journal]{IEEEtran}
\usepackage{amsmath,amsfonts}
\usepackage{algorithmic}
\usepackage{algorithm}
\usepackage{array}
\usepackage[caption=false,font=normalsize,labelfont=sf,textfont=sf]{subfig}
\usepackage{textcomp}
\usepackage{stfloats}
\usepackage{url}
\usepackage{verbatim}

\usepackage{graphicx}
\usepackage{epstopdf}
\usepackage{cite}
\usepackage{amsmath}
\usepackage{mathrsfs}

\begin{document}

\title{Markov Chain-Guided Graph Construction and Sampling Depth Optimization for EEG-Based Mental Disorder Detection}

\author{Yihan Wu, Tao Chang, Peng Xu$^*$, Yangsong Zhang$^*$
\thanks{Yihan Wu and Yangsong Zhang are with the School of Computer Science and Technology, Laboratory for Brain Science and Medical Artificial Intelligence, Southwest University of Science and Technology, Mianyang, 621010, China. Yihan Wu is also with the National Clinical Research Center for Geriatrics, West China Hospital, Sichuan University, Chengdu, China. Yangsong Zhang is also with the Key Laboratory of Testing Technology for Manufacturing Process, Ministry of Education, Southwest University of Science and Technology, Mianyang, 621010, China. Tao Chang is with the Department of Neurosurgery, West China Hospital, Sichuan University, Chengdu, China. Peng Xu is with the MOE Key Laboratory for Neuroinformation, University of Electronic Science and Technology of China, Chengdu 610054, China}
\thanks{Corresponding author1: Yangsong Zhang, email: Email: zhangysacademy@gmail.com. Corresponding author2: Peng Xu, email: xupeng@uestc.edu.cn. }
}



\maketitle

\begin{abstract}
Graph Neural Networks (GNNs) have received considerable attention since its introduction. It has been widely applied in various fields due to its ability to represent graph structured data. However, the application of GNNs is constrained by two main issues. Firstly, the "over-smoothing" problem restricts the use of deeper network structures. Secondly, GNNs' applicability is greatly limited when nodes and edges are not clearly defined and expressed, as is the case with EEG data.In this study, we proposed an innovative approach that harnesses the distinctive properties of the graph structure's Markov Chain to optimize the sampling depth of deep graph convolution networks. We introduced a tailored method for constructing graph structures specifically designed for analyzing EEG data, alongside the development of a vertex-level GNN classification model for precise detection of mental disorders. In order to verify the method's performance, we conduct experiments on two disease datasets using a subject-independent experiment scenario. For the Schizophrenia (SZ) data, our method achieves an average accuracy of 100\% using only the first 300 seconds of data from each subject. Similarly, for Major Depressive Disorder (MDD) data, the method yields average accuracies of over 99\%. These experiments demonstrate the method's ability to effectively distinguish between healthy control (HC) subjects and patients with mental disorders. We believe this method shows great promise for clinical diagnosis.
\end{abstract}


\section{Introduction}
\IEEEPARstart{M}{ental} disorders are characterized by severe disturbances in cognition, emotional regulation, or behavior, affecting approximately 970 million people worldwide in 2019 according to the World Health Organization \cite{WHO}. Common mental disorders include Schizophrenia (SZ), Major Depression Disorders (MDD), Anxiety Disorders, Bipolar Disorder, Post-Traumatic Stress Disorder, among others. Patients with mental disorders often experience a decreased life expectancy of 10-20 years compared to the general population \cite{life1, life2, life3}, yet a large percentage of them have not received effective treatment \cite{WHO, treat1}. One of the major challenges is the lack of significant biological markers \cite{mark1, mark2}.

Various techniques have been used to map human physiological signals, including electroencephalography (EEG), functional magnetic resonance imaging (fMRI), functional near-infrared spectroscopy (fNIRS), and Positron Emission Tomography (PET). Due to advantages such as high temporal resolution, non-invasiveness, low cost, and portability, EEG has become an integral tool for studying neural activity in the brain \cite{EEG1, EEG2, EEG3, EEG4, EEG5}.

Due to the current state where clinically relevant decisions regarding mental disorders are predominantly based on physicians' experiences rather than quantifiable objective indicators, various mental disorder detection methods based on Feature Engineering (FE) have been proposed \cite{review}. These methods typically involve artificial feature extraction, feature selection, and classification.
In the field of SZ detection, Vázquez et al. \cite{backround_SZ1} proposed a method for detecting SZ. They extracted generalized partial directed coherence (GPDC) and direct directed transfer function (dDTF) features and classified them using the random forest (RF) algorithm. They conducted subject-dependent and leave-$p$-subject-out experiments, achieving average area under the curve (AUC) values of 99\% and 87\%, respectively.
Siuly et al. \cite{backround_SZ2} introduced an SZ detection method involving empirical mode decomposition (EMD). EMD decomposes raw EEG signals into intrinsic mode functions (IMFs) and calculates statistical features from these IMFs. They used the Kruskal-Wallis test to select five statistically significant features: Maxima, Minima, Standard Deviation, Activity Natural, and First Quartile (Q1), from the 22 features. By employing Ensemble Bagged Tree (EBT) for classification, they achieved an accuracy of 93.21\% in a 10-fold cross-validation experiment.
In the field of MDD, Mumtaz et al. \cite{backround_MDD1} proposed a method for detecting MDD. They computed alpha interhemispheric asymmetry and spectral power as artificial features and selected the most significant features based on the receiver operating characteristics (ROC) criterion. Finally, they trained a Support Vector Machine (SVM) classifier using these features. In a 10-fold cross-validation experiment, their method achieved a classification accuracy of 98.4\%.
Saeedi et al. \cite{backround_MDD2} introduced a method based on enhanced k-nearest neighbors (E-KNN) for MDD detection. They employed Welch's periodogram method and wavelet packet decomposition to obtain linear features and calculated Approximate Entropy and Sample Entropy as nonlinear features. Using a genetic algorithm (GA) for feature selection, they classified the selected features with an E-KNN classifier. Their method achieved a classification accuracy of 98.44\% in a 10-fold cross-validation experiment.

While the current reliance on artificial feature extraction and selection is highly susceptible to the experience and prior knowledge of researchers, the rapid development of deep learning has addressed this problem to some extent. Deep learning (DL) methods have the capability to perform automatic feature engineering and extract differential components of EEG without manual intervention. Consequently, the field of mental disorder detection has witnessed the emergence of numerous deep learning algorithms in recent years.
In the field of SZ detection, Oh et al. \cite{backround_SZ3} introduced a deep convolutional neural network (CNN) model. They conducted experiments on both subject-dependent and subject-independent scenarios, designing distinct models to achieve the best performance. In the subject-dependent experiment, their model consisted of four convolution layers, five max-pooling layers, and two fully connected layers. In the subject-independent scenario, they employed five convolution layers, two max-pooling layers, two average-pooling layers, one global average pooling layer, and one linear layer. Their models achieved average accuracies of 98.07\% and 81.26\%, respectively.
In 2021, Shoeibi et al. \cite{backround_SZ4} proposed a CNN-LSTM model for SZ detection. They conducted experiments comparing traditional machine learning (ML) methods (SVM, KNN, decision tree, naïve Bayes, RF, extremely randomized trees, and bagging) with DL methods (1D-CNN, long short-term memory (LSTM), and CNN-LSTM). The results showed that the CNN-LSTM architecture achieved the best performance, with an average accuracy of 99.25\% in a 5-fold cross-validation experiment.
In the field of MDD detection, Acharya et al. \cite{backround_MDD3} proposed a Deep CNN method in 2018. They designed a 13-layer CNN model to classify EEG signals acquired from depressed subjects and healthy controls. The model consisted of five CNN layers, five max-pooling layers, and three fully connected layers. They achieved accuracies of 93.54\% and 95.49\% using EEG signals from the left and right hemisphere, respectively. These results indicated that the right hemisphere EEG data is more dominant in depression detection, consistent with the notion that depression is associated with an extremely active right hemisphere.
Song et al. \cite{backround_MDD4} introduced their end-to-end method named LSDD-EEGNet for MDD detection. They combined the efficiency of CNN for feature extraction and the superiority of LSTM for time-series signals in their model. Additionally, they employed a domain discriminator to reduce the discrepancy between training and test datasets by adjusting the data representation space. Through a series of subject-independent experiments, they achieved an average accuracy of 94.69\%, which surpassed traditional ML methods and state-of-the-art methods.

With the emergence of GCN theory, graph neural networks (GNNs) have rapidly found applications in various fields~\cite{TNNLS1, TNNLS2}. Graphs are typical non-Euclidean data structures that represent the content and structural features of graph data using nodes (vertices) and relationships between nodes (edges). Graph structures, such as molecular structure diagrams and social networks, contain richer information than Euclidean structures like time series and images, even though they lack features such as translational invariance due to the variable number of neighboring nodes. Given the presence of links between nodes, graph data structures present an opportunity to capture intricate relationships. Spectral-based graph convolution algorithms, represented by GCN, and spatial-based graph convolution algorithms, represented by GraphSAGE, are common graph neural networks. GraphSAGE samples a fixed number of neighbors and learns a set of aggregators to combine the obtained information for the representation of the current node. However, existing graph convolutional neural networks face limitations due to over-smoothing issues, which hinder the utilization of deep networks for feature extraction. In this paper, we propose a novel method for organizing graph data structures by calculating and analyzing their structural features based on the theory of Markov Chain. We develop a deep graph convolutional network called DeepSAGE, based on the GraphSAGE algorithm, for the classification of two mental disorders. To tackle the over-smoothing problem, we introduce an indicator to measure the severity of over-smoothing based on the theory of Markov Chain. Experimental results demonstrate that DeepSAGE achieves an accuracy of 100\% and 99.9\% on SZ and MDD datasets, respectively.

The primary contribution of this article is:
\begin{itemize}
	\item [$\bullet$] Proposed a novel EEG signal graph organization method for detecting mental disorder. 
	\item [$\bullet$] Proposed a high-accuracy deep graph convolution method for classifying graph data.
	\item [$\bullet$] Proposed a method for optimizing the graph structure and corresponding sampling depth of deep graph convolution networks.
\end{itemize}

The remainder of this paper is organized as follows. Section 2 introduces materials and methods. Section 3 describes the settings and results of extensive experiments, comparisons between MSBAM and the baseline methods are also be provided in this section. Section 4 and 5 present the discussions and conclusion.

\section{Materials and Methods}
\subsection{Datasets}
Two public dataset of SZ and MDD were adopted to validate the performance of our methods, which are widely used in research for detecting these two mental disorders.

\textbf{For SZ}: This dataset was published by the Institute of Psychiatry and Neurology in Warsaw, Poland. It includes EEG recordings from 14 SZ patients (7 males: 27.9 $\pm$ 3.3 years, 7 females: 28.3 $\pm$ 4.1 years) and 14 healthy controls (7 males: 26.8 $\pm$ 2.9, 7 females: 28.7 $\pm$ 3.4 years). The EEG signals were recorded while the participants' eyes were closed for a period of 15 minutes, with a sampling rate of 250Hz. Nineteen electrodes, including Fp1, Fp2, F7, F3, Fz, F4, F8, C3, Cz, C4, T3, T4, T5, T6, P3, Pz, P4, O1, and O2, were used according to the international standard 10-20 system. More details can be found in the reference~\cite{Dataset_SZ}.

\textbf{For MDD}: This public dataset was published by Mumtaz et al.~\cite{Dataset_MDD}. It consists of EEG recordings from 34 MDD patients (17 males and 17 females, 40.3 $\pm$ 12.9 years) and 30 HCs (21 males and 9 females, 38.3 $\pm$ 15.6 years). Data from 30 MDD patients and 28 HCs were provided. The recordings were obtained during a 5-minute eye-closed resting-state EEG session, with a sampling rate of 256 Hz. Like the SZ dataset, the electrode placement follows the international 10-20 system, and nineteen electrodes were utilized in this study. Further information can be found in the original paper~\cite{Dataset_MDD}."

\textbf{For MDD}: This public dataset was published by Mumtaz et al.~\cite{Dataset_MDD}. This dataset consists of EEG signals recording from 34 MDD patients (17 males and 17 females, 40.3 $\pm$ 12.9 years) and 30 HC (21 males and 9 females, 38.3 $\pm$ 15.6 years). Of these, data from 30 MDD patients and 28 HC were provided. The eye-closed resting-state EEG data were collected for 5 minutes. Nineteen electrodes were utilized with a sampling rate of 256 Hz. Similar to the SZ dataset, the electrodes are placed following the international 10-20 system. More information can be found in the original paper~\cite{Dataset_MDD}.
	
\subsection{Data pre-processing}

We used a similar pre-processing procedure on both datasets. First, to enhance the signal to noise ratio, we empirically employed a bandpass filter with a frequency of 0.5-48 Hz for the SZ dataset and 0.5-70 Hz for the MDD dataset. The data will then be segmented by a 5-second slide window. Additionally, we employed a threshold filter to remove the ocular artifacts by dropping the segment which amplitudes value is out of [-100, 100]$\mu$V on the MDD dataset. Finally, common reference is applied to obtain the processed data.

\subsection{Graph organization}
Due to the fact that the input data in this study is multi-channel EEG signals rather than natural graph structures, it is necessary to manually construct a graph structure for the signals. The method of graph construction is one of the most important factors affecting the performance of graph neural network methods. In this paper, a novel graph structure is proposed to organize all samples of the target subjects. In the graph, each node is used to represent the information of an individual sample, and artificially constructed edges are used to ensure information flow during GNN sampling.

In conventional graph structures, the "over-smoothing" phenomenon often occurs during high-order sampling due to the high connectivity between nodes. The "over-smoothing" phenomenon refers to the convergence of the expression of each node to a fixed subspace as the number of graph convolutional layers increases. This phenomenon leads to uniform expression across nodes, which results in loss of differentiated information and makes it no longer linearly separable in Euclidean space.

The research introduced by Xu et al.~\cite{GraphXu} has demonstrated that if a central node is chosen as the starting point for random walks on a graph, the range of the walk can quickly cover the entire graph. On the other hand, if a peripheral node is selected as the starting point, it will greatly slow down the coverage of the entire graph. Inspired by this, we designed the graph structure according to the following principles:
\begin{itemize}
\item [$\bullet$] Maximizing the placement of nodes in the peripheral region of the graph: By strategically positioning a greater proportion of nodes in the periphery, we aim to enhance the adaptability of data to deep neural networks while mitigating the adverse effects of over-smoothing. This approach helps preserve the distinctive characteristics of individual samples and promotes better discriminability within the graph structure.

\item [$\bullet$] Ensuring structural similarity among data from the same subject: We recognize the importance of maintaining structural consistency for samples originating from a specific subject. By enforcing this principle, we establish a foundation for fair comparisons and facilitate meaningful analysis across different samples. It enables us to capture subject-specific patterns and unveil underlying relationships in a more accurate and reliable manner.

\item [$\bullet$] Controlling the number of edges in the graph: Redundant connections within the graph can impede the flow of information and introduce noise in subsequent analyses. Therefore, we exercise careful control over the number of edges, striking a balance between capturing essential relationships and minimizing the presence of superfluous connections. This optimization helps streamline the information flow and enhances the efficiency of subsequent graph-based analyses.
\end{itemize}

Based on the aforementioned guiding principles, we propose a novel graph structure to organize the samples obtained from all subjects under study. In this graph representation, each node corresponds to the information encapsulated within an individual sample. The edges connecting these nodes are manually constructed to ensure seamless information circulation during sampling with GNNs. By adopting this well-defined graph structure, we can harness the power of GNNs effectively and uncover intricate patterns and connections within the multi-channel EEG signals.

\begin{figure}[!htb]
	\centering
	\subfloat[]{\includegraphics[scale=0.20]{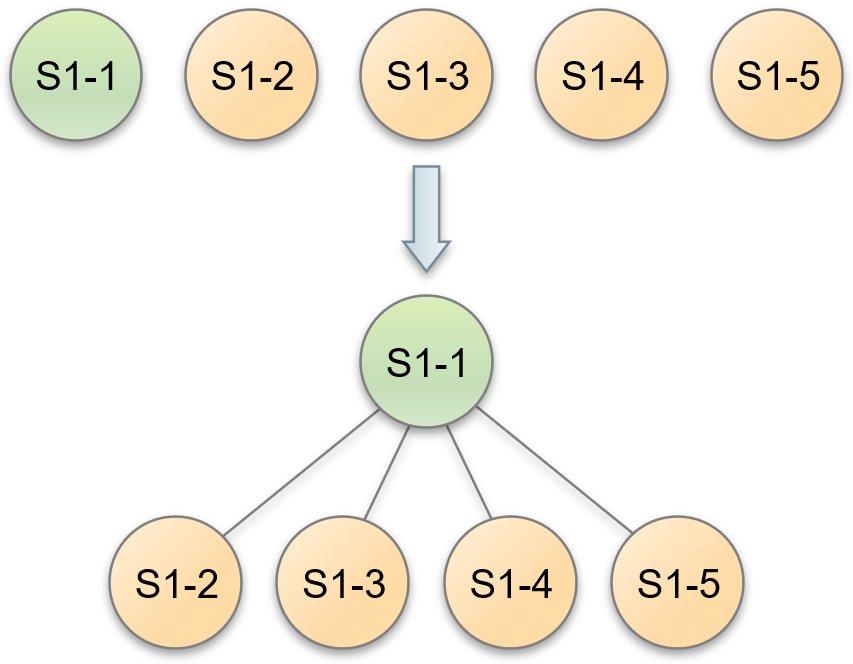}
		\label{graph_subgraph}}
	\hfil
	\subfloat[]{\includegraphics[scale=0.24]{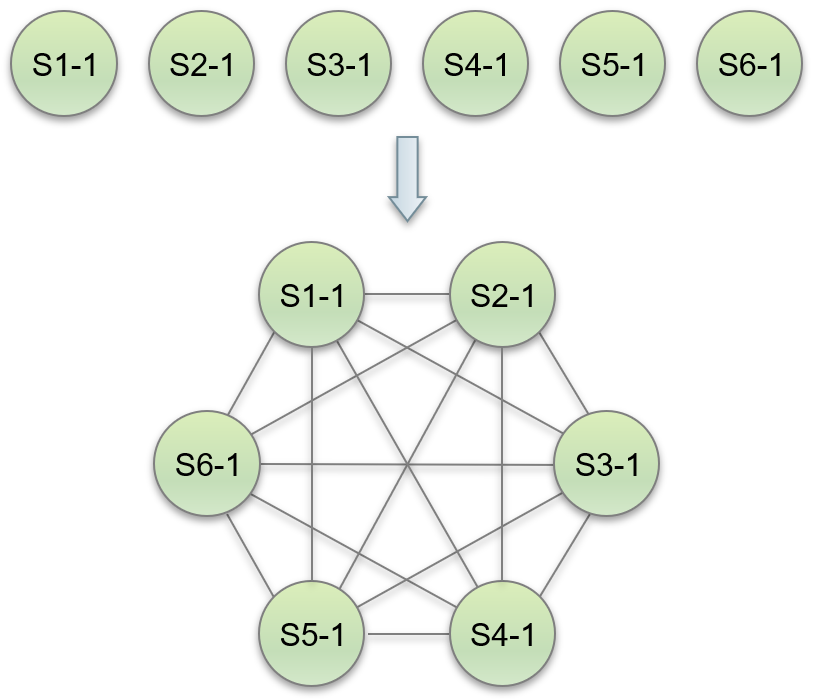}
		\label{graph_wholegraph}}
	\caption{Overview of Graph Data Construction
a) Example of Subgraph Construction for Subject No. 1
This panel shows the process of constructing a subgraph for a single subject, labeled as No. 1, with five available EEG samples. One of these samples, labeled as S1-1, is arbitrarily chosen to serve as the root vertex. The remaining samples are then connected to S1-1 via undirected edges, creating a subgraph specific to Subject No. 1. b) Method of Connecting Subgraphs with Undirected Edges
This panel illustrates the overall strategy for connecting all of the subgraphs together using undirected edges. Each green circle denotes a subgraph corresponding to an individual subject, as depicted in Figure 1a). To establish inter-subject connectivity, undirected edges are added between the root vertices of every pair of subgraphs.
}
	\label{graph}
\end{figure}

For each participant in the study, the EEG samples ($S\in\mathbb{R}^{C \times T} $) undergo an initial transformation into one-dimensional vectors ($S\in\mathbb{R}^{(C * T)} $). These transformed vectors can be regarded as the features associated with the corresponding nodes in the graph. Within each subject, a single sample is arbitrarily selected to serve as the \emph{root vertex} of the subgraph. Subsequently, an undirected edge is established between each chosen \emph{leaf vertex} and its respective \emph{root vertex}, effectively constructing a subgraph that is specifically associated with that subject, as visually represented in Figure~\ref{graph_subgraph}. This subgraph structure plays a crucial role in ensuring a high level of cohesiveness among the graph structure features for all nodes within each subject. It promotes the identification and understanding of subject-specific patterns and relationships within the EEG data.

To ensure connectivity among vertices from different subjects, the root vertices of all subgraphs were interconnected by undirected edges, forming a complete graph. This construction approach guarantees communication between vertices across subjects, facilitating information exchange and analysis. For an example, the process of organizing subgraphs using data from six subjects is illustrated in Figure 1. In Figure 1, the root vertices of the corresponding subject subgraphs are labeled as S1-1, S2-1, and so on. This organization enables efficient spatial domain sampling utilizing the SAGEConv operator. By applying this sampling technique, the graph exhibits a high level of cohesiveness within the subject data. Furthermore, it enables information exchange not only among leaf vertices within each subgraph but also between leaf vertices residing in different subgraphs. This exchange occurs through a network of interconnected root vertices, thereby minimizing redundant connections and optimizing the flow of information. Through the use of this interconnected graph structure, researchers can effectively analyze and explore the subject data with improved connectivity and reduced redundancy.

\subsection{Network structure}

\subsubsection{GCN Convolution}
GCN is one of the most classical frequency domain-based transductive Graph Neural Network methods. It utilizes the generalization of Discrete Fourier Transformation and convolution theorem on graphs to implement graph operations in the frequency domain. The Graph Fourier Transform (GFT) of a graph X is described as follows:
\begin{eqnarray}
	&&\mathbf{\hat{X}}=F(\mathbf{X})=\mathbf{U^TX}\\
	&&\mathbf{L=U^T\Lambda U} \label{GFT}
\end{eqnarray}
where L denotes the Laplacian matrix of the graph, and F denotes GFT. 
According to the convolution theorem, the convolution operation on a graph X using a convolution kernel H can be expressed as:
\begin{eqnarray}
	&\mathbf{(X * H)}_G&=F^{-1}[F(\mathbf{X})\cdot F(\mathbf{H})]\\
	&&\mathbf{=U(U^TX\cdot \hat{H})}\\
	&&\mathbf{=U(} diag(\hat{g}_0, \hat{g}_1, ..., \hat{g}_n)\mathbf{U^TX}\\
	&&\mathbf{=U(}diag(\theta_0, \theta_1, ..., \theta_n)\mathbf{U^TX}\\
	&&\mathbf{=U\Theta U^TX} \label{GraphConv}
\end{eqnarray}
The final GCN formula can be obtained by utilizing the first-order approximation of Chebyshev polynomials as follows:
\begin{eqnarray}
	&&\mathbf{(X * H)}_G = \mathbf{\widetilde{D}}^{-\frac{1}{2}} \mathbf{\widetilde{A}} \mathbf{\widetilde{D}}^{- \frac{1}{2}} \mathbf{X} \mathbf{\Theta} \label{Cheb}
\end{eqnarray}
where $\widetilde{D}$ and $\widetilde{A}$ are the degree matrix and adjacency matrix with self-connections added, respectively. W is a trainable parameter in this context.

\subsubsection{SAGE Convolution}
GraphSAGE is a classical inductive graph convolution method based on the spatial domain\cite{SAGE}. Unlike the GCN algorithm, which is a frequency domain-based transductive algorithm, GraphSAGE directly samples the local neighboring vertices of the target vertex in the spatial domain. Additionally, GraphSAGE utilizes inductive sampling algorithms to generalize the model to unseen vertices\cite{SAGE}. In contrast to Euclidean space, where each data point has a fixed number of neighbors, graph data can have a varying number of neighbors for each vertex. Spectral-based GCN requires sampling all neighbors of each node across the entire graph to learn embedding representations for each vertex. However, GraphSAGE takes a different approach by randomly sampling a fixed number of neighbors for each vertex in the spatial domain, regardless of the number of neighbors, and then uses learned aggregators to combine the sampled information and obtain the representation of the current vertex.

When the differentiable aggregator function is set to Mean function, the core operator of the GraphSAGE, termed as SAGEConv, is able to be described as follows:
\begin{equation}
	\mathbf{x}_{i+1} = \mathbf{W}_1 \mathbf{x}_i + \mathbf{W}_2 \cdot \mathrm{mean}_{j \in \mathscr{N(i)}} \mathbf{x}_j
\end{equation}
where $x_i$ denotes the representations of the vertex in $i$-th layer of the network. The $\mathrm{mean}_{j \in \mathscr{N(i)}} \mathbf{x}_j$ described the mean features of the node's local neighborhood.

\subsubsection{DeepSAGE}
\begin{figure*}[ht]
	\centering
	\includegraphics[scale=0.33]{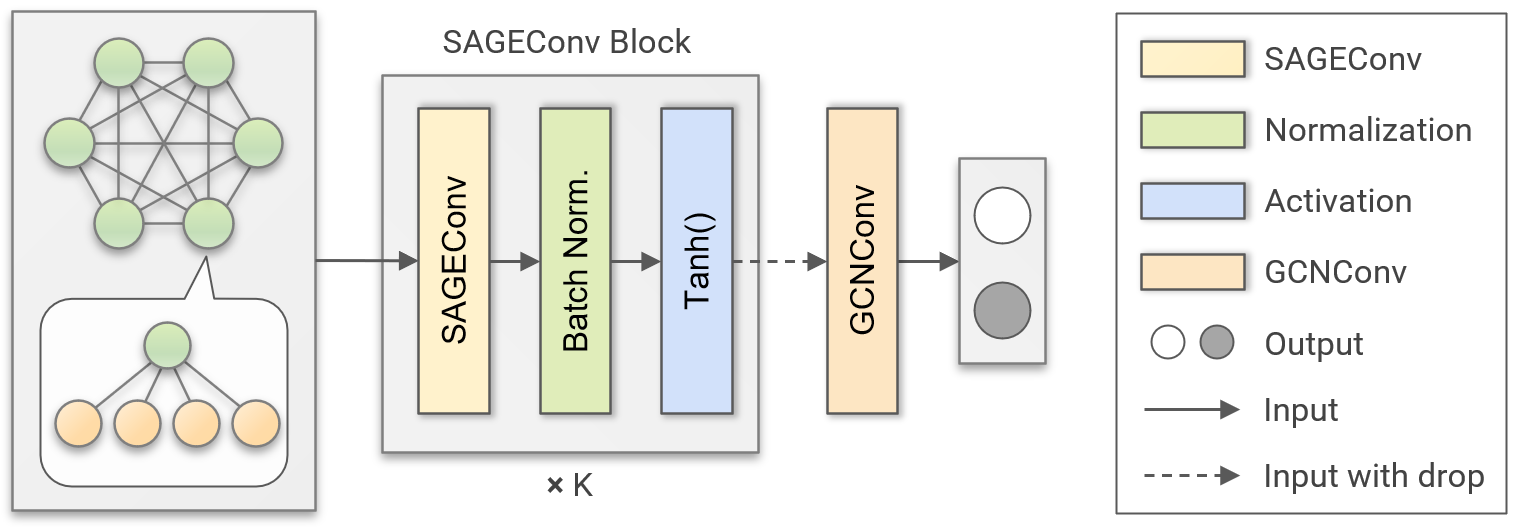}
	\caption{The structure of DeepSAGE. The SAGEConv block are stacked \emph{K} times for extract the higher-level abstract feature of the input graph.}
	\label{DeepSAGE}
\end{figure*}

For the graph data introduced in previous, we built a deep graph neural networks to extract the higher-level abstract representations of the vertices. The proposed method is named as DeepSAGE, which is consist of multiple SAGEConv blocks and a GCNConv classifier. The structure of DeepSAGE is shown in Fig.~\ref{DeepSAGE}.

When the graph data, denoted as $\mathcal{G\{V, E\}}$, was input into the DeepSAGE, it firstly was feed into \emph{K} concatenated SAGEConv block. Each block is consisted of one SAGEConv layer, one Batch Normalization layer and one Tanh as the nonlinear activation function, which is illustrated in the following:

To extract higher-level abstract representations of the vertices from the graph data mentioned previously, we developed a deep graph neural network called DeepSAGE. DeepSAGE comprises multiple SAGEConv blocks and a GCNConv classifier. The architecture of DeepSAGE is depicted in Figure 2.

When the graph data, represented as $\mathcal{G\{V, E\}}$ , is fed into DeepSAGE, it undergoes a series of operations. Firstly, the data is passed through \emph{K} concatenated SAGEConv blocks. Each block consists of a SAGEConv layer, a Batch Normalization layer, and a Tanh activation function, as illustrated below:
\begin{eqnarray}
	&&h_0 = \mathcal{G\{V, E\}}\\
	&&z_k = SAGEConv(h_k)\\
	&&h_{k+1} = Tanh(BN(z_k)) \label{SAGEConv block}
\end{eqnarray}
After undergoing \emph{K} iterations of the SAGEConv block, the high-level representations are fed into a GCNConv classifier. Prior to this, a dropout layer is applied with a rate of 0.5 to prevent overfitting. The GCNConv layer generates two features for each vertex, which represent the classification scores. These scores are positively correlated with the probabilities ($P(c|G), c= 0,1 $) of the vertex belonging to the two classes. This relationship can be described as follows:
\begin{eqnarray}
	&&\hat{y}^i= \mathop{\arg\max} Cls(h_n^i) \label{Predict}
\end{eqnarray}
where $\hat{y}^i$ denotes the predicted label of $i$-th vertex, and the $Cls$ denotes the GCNConv classifier.

The GCNConv layer processes the high-level representations to produce the scores indicating the likelihood of a vertex belonging to each class. A higher score for Class 1 suggests a higher probability of membership in that class, while a lower score indicates a higher likelihood of belonging to Class 0. 

The value of \emph{K} is critical to determine if a model can effectively work. A small value for \emph{K} would make it challenging to capture higher-level feature representations, while an excessively large value would result in the issue of "Over-smoothing". In this study, we aim to leverage the properties of Markov chains to calculate the appropriate sampling depth for the input graph structure under multiple random samplings.

\subsubsection{Graph structure analysis}
When conducting \emph{K}-order deep random sampling using SAGEConv, the fundamental essence of its feature source can be likened to performing multiple random walks with a depth of \emph{K}. As the number of layers increases, the receptive field of an individual vertex gradually expands as well.
To ensure a reasonable sampling depth, this study adopts a random walk technique on graph data, simulating multiple samplings with an arbitrary vertex selected as the starting position. The objective is to observe the information sources at each layer of sampling. To facilitate analysis, all vertices in the entire graph are abstracted into four groups. Group $L$ represents leaf vertices that belong to the same subgraph as the initial vertex. Given their structural similarity within the subgraph, these leaf vertices are divided into a single category. Group $R$ corresponds to the root vertex associated with the initial vertex. Group $\hat{R}$ and $\hat{L}$ respectively encompass the root vertices and leaf vertices from other subgraphs. For the initial vertex, all root vertices from other subgraphs share the same structural characteristics, including topology and connectivity, but may have unknown features and labels. Therefore, they are classified into the same group.

\begin{figure*}[ht]
	\centering
	\includegraphics[scale=0.25]{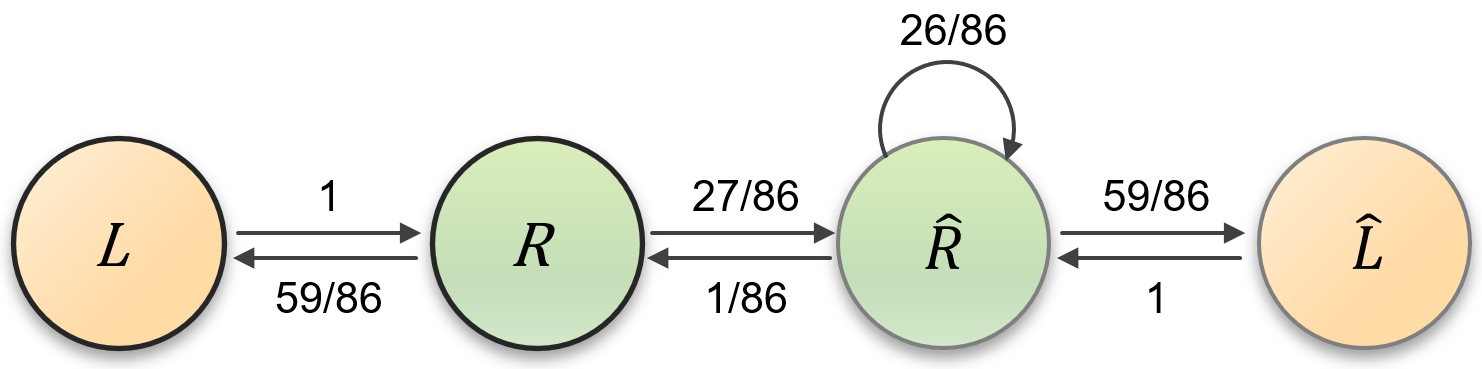}
	\caption{The Markov model of SZ dataset. The dataset has 28 subjects and 60 samples for each one.}
	\label{Markov model}
\end{figure*}

Through analyzing the network topology of the input dataset, we are able to calculate the initial states and state transition matrix for every sampling event, which serves as the basis for constructing a comprehensive Markov chain model. To demonstrate the proposed methodology efficiently, we have implemented it to the SZ dataset, and the resulting model is depicted in Figure~\ref{Markov model}. 
For arbitrary initial vertex in this graph, we construct its initial state probability matrix in the order of "$L$-$R$-$\hat{R}$-$\hat{L}$" as follows:
\begin{eqnarray}
	&&\begin{bmatrix}
		\dfrac{59}{60} & \dfrac{1}{60} & 0 & 0
	\end{bmatrix}\nonumber
\end{eqnarray}
and the corresponding probability transition matrix is given by:
\begingroup
\renewcommand*{\arraystretch}{2.1}
\begin{eqnarray}
	&&\begin{bmatrix}
		0 & 1 & 0 & 0\\
		\dfrac{59}{86} & 0 & \dfrac{27}{86} & 0\\
		0 & \dfrac{1}{86} & \dfrac{26}{86} & \dfrac{59}{86}\\
		0 & 0 & 1 & 0
	\end{bmatrix}\nonumber
\end{eqnarray}
\endgroup

In the previously described Markov model, let $S_0$ represent the initial state probability and $P$ denote the state transition probability matrix. Thus, the probability of the state for the $k$-th sampling can be expressed as follows:
\begin{equation}
	\mathbf{X_k} = \mathbf{X_{k-1}P} 
\end{equation}

As a result of the properties exhibited by Markov chains, as the depth of sampling, represented by $K$, increases, the probabilities associated with each state will converge towards a fixed value. This convergence is determined by the transition probability matrix, $P$, and is independent of the initial state, denoted as $X_0$.
In the previously mentioned Markov model of the SZ dataset, we generated a line chart to visually represent the probabilities of sampling each category at the $K$-th layer.
\begin{figure*}[!htb]
	\centering
	\includegraphics[scale=0.3]{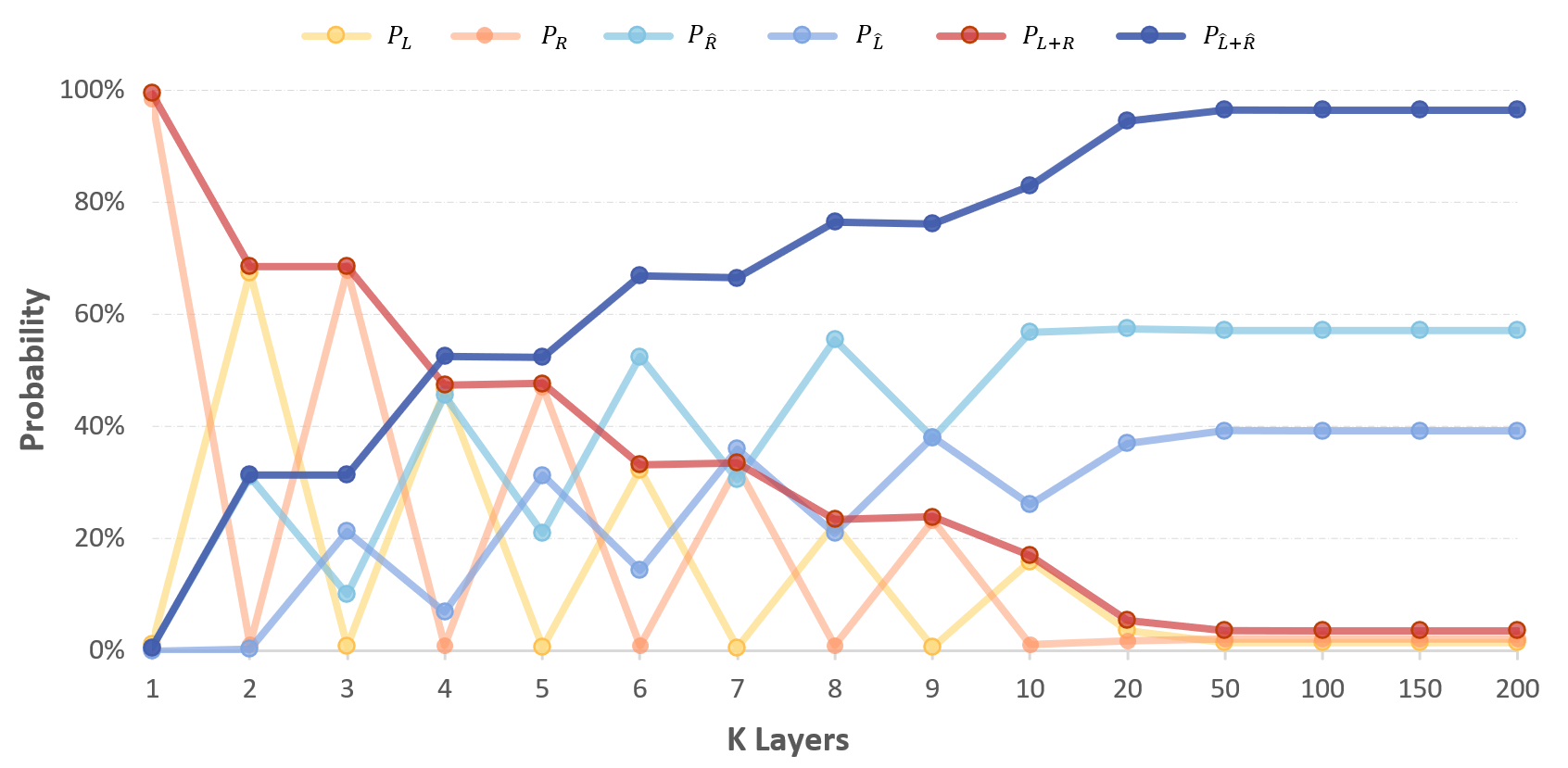}
	\caption{Probabilities of sampling each category at the $K$-th layer. When $K$>50, the probability of sampling has already reached convergence with respect to this stationary distribution}
	\label{Structure}
\end{figure*}

Our study observed that when $K$ exceeds 50, the probability of sampling has already reached convergence with respect to this stationary distribution. This phenomenon is known as "over-smoothing" in graph neural networks. However, due to the specific graph structure we designed, the convergence speed on this graph data is relatively slow, which enables us to employ deeper Graph Convolutional Networks to learn expressive features. In this paper, we select the $P_{L+R}$ value, which is determined by the ratio of the number of subjects to the sample size of a single subject, to quantify the compatibility between the depth of network sampling ($K$) and the structure of the graph network. To avoid insufficient model learning ability caused by excessively large values of $P_{L+R}$ or the "over-smoothing" problem caused by overly small values, it is crucial to maintain a reasonable range for $P_{L+R}$ that can balance inter-subject diversity and sampling richness. 

In the upcoming experiments, we will rely on this indicator, $P_{L+R}$, to guide the adjustment of the sampling depth of DeepSAGE and explore its optimal range of values.

\section{Experiments and Results}
\subsection{Model implementation}
In consideration of the $P_{L+R}$ indicator, we ultimately selected the following parameters for our model.

To begin with, we utilized 5 layers of SAGEConv for graph sampling, with an output dimension of 128 for each SAGEConv layer. For optimization, we employed the Adam optimizer with a learning rate of 3e-5, specifically targeting the minimization of the cross-entropy loss function.

In the evaluation phase, we adopted the leave-one-subject-out (LOSO) approach and reported the mean results across all subjects, together with the subject-level standard deviation.

Additionally, we conducted multiple experiments on both the SZ and MDD datasets, calculating the average accuracy, sensitivity and specificity. To ascertain the effectiveness of our approach, we also compared the obtained results with those of the current baseline model.
Furthermore, we conducted experiments and analysis on the impact of different values of $P_{L+R}$ on model performance, demonstrating that excessively high or low values of $P_{L+R}$ do indeed affect the model's performance. Our research findings emphasize the importance of selecting appropriate values of $P_{L+R}$ for optimizing model performance, providing valuable insights for further improvements and applications.

\subsection{Results of SZ}
In the case of the SZ dataset, as each individual subject's data had a long length, we selected the first 60 samples from each subject's data to form the experimental dataset. The performance results are presented in Table~\ref{SZ result}.
\begin{table}[htb]
	\centering
	\renewcommand{\arraystretch}{1.15}
	\setlength{\tabcolsep}{7pt}
	\caption{Classification results of DeepSAGE and all compared methods on the SZ dataset. The ACC, SEN and SPE in the title denotes accuracy (mean$\pm$std), sensitivity and specificity, respectively.}
	\scalebox{1}{\begin{tabular}{lccc}
			\hline
			\hline
			Methods  & ACC(\%) & SEN(\%) & SPE(\%)  \\
			\hline
			DeepConvNet~\cite{RAE} & 58.96$\pm$6.92 & 60.24 & 55.33\\
			DCNN~\cite{RAE}       & 73.21$\pm$4.74  & 71.91 & 75.18  \\
			EEGNet~\cite{RAE}       & 77.18$\pm$0.96  & 74.58 & 79.36  \\
			RAE~\cite{RAE}       & 81.81$\pm$1.60  & 80.30 & 83.37  \\
			\textbf{DeepSAGE}       & \textbf{$ \sim $100.00$\pm$0.00}  & \textbf{$ \sim $100} & \textbf{$ \sim $100}  \\
			\hline
			\hline
	\end{tabular}}
	\label{SZ result}
\end{table}

To validate the model performance, we conducted five repeated experiments, and achieved the accuracies of [100\% 100\%, 100\%, 100\%, 99.94\%]. The experimental results demonstrated that the DeepSAGE has superior performance and stability.

%
To analyze the impact of the $P_{L+R}$ values, we conducted experiments on each subject with different data lengths and collected performance metrics. We compared the $P_{L+R}$ values of the current network parameters to assess their influence. To ensure consistent evaluation, we set the depth to a fixed value of K=5 in subsequent experiments without modifying the network structure. The 'All' column in the displayed table represents the average $P_{L+R}$ value calculated by considering the variations in data length across subjects. Additionally, we fine-tuned the hyper-parameters to achieve optimal results when using all available samples. Please refer to Table~\ref{SZ result_nSample} for detailed results.

\begin{table}[htb]
	\centering
	\renewcommand{\arraystretch}{1.15}
	\setlength{\tabcolsep}{7pt}
	\caption{The accuracy and $P_{L+R}$ values under different sample slices on the SZ dataset. For the last row, we calculated the $P_{L+R}$ value based on the average sample size across all subject.}
	\scalebox{1}{\begin{tabular}{lcc}
			\hline
			\hline
			Samples & ACC(\%)  & $P_{L+R}$(\%)\\
			\hline
			10 & 96.43$\pm$18.90  & 8.85 \\
			\textbf{60} & \textbf{$ \sim $100.00$\pm$0.00}   & \textbf{47.68}\\
			All & 74.86$\pm$42.67  & 78.17 \\
			\hline
			\hline
	\end{tabular}}
	\label{SZ result_nSample}
\end{table}

The experimental results demonstrated that excessively high or low $P_{L+R}$ values may affect recognition accuracy. In actual experiments, the $P_{L+R}$ value can be adjusted to an appropriate range by adjusting the number of subjects used and the sample size per subject.

\subsection{Results of MDD}
In contrast to the SZ dataset, the MDD dataset consists of 58 participants, each with an average of only 55 samples. We conducted two sets of experiments to analyze the dataset. In the experiment, the entire dataset was evenly divided into two parts, creating two minor datasets. To ensure label balance, half of the 30 MDD and 28 HC participants were assigned to each part, respectively. The final result was obtained by calculating the average results of the two minor datasets. Table~\ref{MDD result} presents the average accuracy results from both sets of experiments, comparing them with the baseline results.
\begin{table}[htb]
	\centering
	\renewcommand{\arraystretch}{1.15}
	\setlength{\tabcolsep}{7pt}
	\caption{The best classification results of DeepSAGE and all compared methods on the MDD dataset. The ACC, SEN and SPE in the title denotes accuracy (mean$\pm$std), sensitivity and specificity, respectively.}
	\scalebox{1}{\begin{tabular}{lccc}
			\hline
			\hline
			Methods  & ACC(\%) & SEN(\%) & SPE(\%)  \\
			\hline
			EEGNet~\cite{XiaMin} & 88.69$\pm$ 1.44 & 88.12$\pm$ 1.54 & 88.86$\pm$ 2.29\\
			DeprNet~\cite{XiaMin}       & 88.54$\pm$ 0.77 & 87.29$\pm$ 1.65 & 88.51$\pm$ 1.07\\
			DSNet~\cite{XiaMin}       &91.69$\pm$ 0.45 & 92.11$\pm$ 0.58 & 90.54$\pm$ 0.63 \\
			\textbf{DeepSAGE}       & \textbf{99.90$\pm$0.23}  & \textbf{99.94} & \textbf{100}  \\
			\hline
			\hline
	\end{tabular}}
	\label{MDD result}
\end{table}

The experimental results demonstrated that the DeepSAGE presented significantly better performance than other methods. To verify the stability of the model, we repeated the experiment five times, following the same protocol as that used for the SZ dataset. The experimental results were [99.90\%, 98.57\%, 98.28\%, 99.72\%, 98.28\%], which will be further analyzed in the Discussion section.
%

Furthermore, to verify the necessity of dividing subjects into two groups to control for graph structure, we compared the accuracy and $P_{L+R}$ values between the second experiment using the complete dataset and the first experiment in which participants were divided into two groups. The results is shown in Table~\ref{MDD result_nGroup}.
\begin{table}[htb]
	\centering
	\renewcommand{\arraystretch}{1.15}
	\setlength{\tabcolsep}{7pt}
	\caption{The accuracy and $P_{L+R}$ value comparison between using the entire dataset, i.e., Group 1, and averaging the two minor datasets, Group 2, in the previous experiment on the MDD dataset.}
	\scalebox{1}{\begin{tabular}{lcc}
			\hline
			\hline
			Group & ACC(\%) & $P_{L+R}$(\%) \\
			\hline
			
			1& 96.12$\pm$14.03  & 24.35 \\				\textbf{2}   & \textbf{100.00$\pm$0.00}   & \textbf{44.06}\\
			\hline
			\hline
	\end{tabular}}
	\label{MDD result_nGroup}
\end{table}

The experimental results demonstrate that the accuracy is significantly higher after subject division compared to before. This phenomenon further verified the hypothesis that maintaining the $P_{L+R}$ value within an optimal  range can lead to improved accuracy in model classification. 

Based on these experiments, we empirically conclude that a $P_{L+R}$ value of around 40\% is appropriate for detecting mental disorders.

\section{Discussion}
In the preceding section, we employed a novel method for organizing graph structures and utilized a corresponding deep graph network for patient detection in two mental disorder datasets. By prioritizing the $P_{L+R}$ values and adapting the graph structures accordingly, we achieved an accuracy rate of nearly 100\%. To further examine the influence of graph structure compatibility with network sampling depth (measured by the $P_{L+R}$ value) on classification accuracy, we conducted a focused analysis on the results with lower accuracy in the MDD dataset.

\begin{figure}[htb]
	\centering
	\includegraphics[scale=0.45]{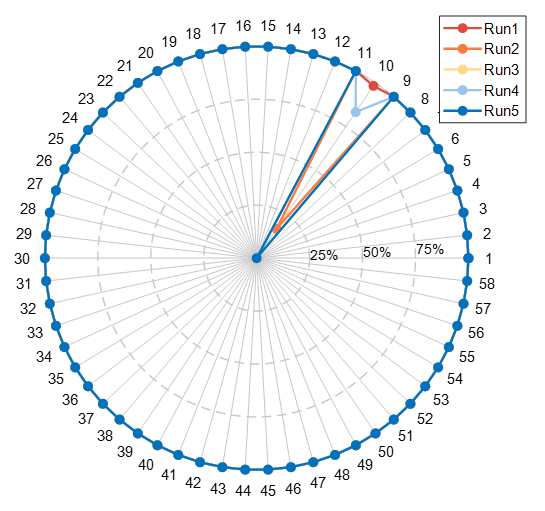}
	\caption{The accuracy of each subject in the MDD dataset across five rounds of experiments.}
	\label{Radar}
\end{figure}

As mentioned earlier, we performed five repeated experiments on the MDD dataset and found that the results lacked sufficient stability. We calculated the accuracy rates for each subject, and these results are shown in Fig.~\ref{Radar}.

It is evident that all subjects, except for Subject 10, achieved 100\% accuracy. We noticed that this might be due to the fact that Subject 10 had only 6 samples after removing ocular artifacts. Therefore, the sampling probability of Subject 10 is significantly different from that of other subjects, resulting in a $P_{L+R}$ value of only 3.42\%. This also supported that an excessively low $P_{L+R}$ value can cause the feature representation of samples to become similar, thereby losing their specificity and resulting in a significant decrease in identification accuracy.

It is evident that all subjects, except for Subject 10, achieved 100\% accuracy. We noticed that this discrepancy may be due to the fact that Subject 10 had only 6 samples remaining after removing ocular artifacts. Consequently, the sampling probability of Subject 10 deviated significantly from that of the other subjects, resulting in a $P_{L+R}$ value of only 3.42\%. This finding further supports the notion that an excessively low $P_{L+R}$ value can cause the feature representation of samples to become similar, leading to a loss of specificity and a significant decrease in identification accuracy.

\section{Conclusion}
In this paper, we proposed a novel EEG signal graph organization method for detecting mental disorder, and  introduced an indicator, namely $P_{L+R}$ value, to measure the degree of matching between the graph structure and the corresponding sampling depth of deep graph convolution networks. We designed a high-accuracy semi-supervised deep graph convolution method for classifying the EEG data guided by the $P_{L+R}$ indicator, named DeepSAGE. We conducted a great number of experiments to verify the performance of the proposed method. After optimizing the detecting method by adjusting the $P_{L+R}$ value, we achieved accuracy rates of 100\% and 99.90\% on the SZ and MDD datasets, respectively.

In this paper, we proposed a novel method for organizing EEG signal graphs to detect mental disorders, accompanied by the introduction of an indicator known as the $P_{L+R}$ value. This indicator allows us to quantify the correspondence between the graph structure and the corresponding sampling depth of deep graph convolution networks. We devised a semi-supervised deep graph convolution method, named DeepSAGE, which leverages the $P_{L+R}$ indicator to guide the classification of EEG data. We conducted an extensive series of experiments to evaluate the performance of the proposed method. By optimizing the detection method through $P_{L+R}$ value adjustments, we achieved remarkable accuracy rates of 100\% and 99.90\% on the SZ and MDD datasets, respectively.

In our future work, we will continue to explore the application of DeepSAGE-related techniques for detecting other types of mental disorders. Additionally, we intend to conduct more extensive research into the influence of the $P_{L+R}$  value on deep graph convolutional networks.

\bibliographystyle{IEEEtran}
\bibliography{citelist}

\vfill

\end{document}